\documentclass[journal]{IEEEtran}

\usepackage{amsmath}
\usepackage{mathtools}
\usepackage{cuted}

\usepackage{graphicx}
\usepackage{adjustbox}
\usepackage{caption}
\usepackage{subcaption}
\usepackage{multirow}
\usepackage{epstopdf}

\title{Optimized Sharing of Coefficients in Parallel Filter Banks}
\author{M. Tun\c{c} Arslan*$^{,\Phi }$, Onur Yorulmaz*, L. Erdin\c{c} At{\i}lgan*\\ 
	*Meteksan Defense Inc., Ankara, Turkey.\\  
	$^{\Phi }$Dept. of Electr. \& Electron. Eng., Bilkent University, Ankara, Turkey. \\
	Email: {tarslan@ee.bilkent.edu.tr, oyorulmaz@meteksan.com, latilgan@meteksan.com} \vspace{-20pt}
}

\begin{document}

\maketitle

\begin{abstract}
Filters are the basic and most important blocks of most signal processing applications. In many applications, a group of parallel  filters are used as filter banks. Parallel filter banks naturally require much more computations. Especially on chip applications, the resources are limited and shared among many algorithms. For this purpose, many filter optimization schemes are proposed to reduce the number of resources that filtering operations require. In this work, a novel optimization algorithm is proposed to decrease the number of operations in a group of parallel filters. The filter coefficients are grouped in a two stage process which enables increased coefficient sharing between different filters. The algorithm is capable of decreasing the number of registers, look-up tables and DSP48s by up to 50\% of a regular parallel filter bank, without requiring increased sampling rate.
\end{abstract}

\section{Introduction}
\label{sec:intro}
Parallel filter banks consists of several filters that modify the input signal in order to extract information. They are widely used in signal processing applications, such as radar signal processing for target detection, communications for synchronization and matched receiver, image processing and convolutional neural networks. 

In radar signal processing applications, matched filter banks are used for both wave compression and as optimum receivers to find the delay and the Doppler shift \cite{richards2005fundamentals}. In radar systems, having higher bandwidth signals are almost always beneficial for better resolution, but this increases the number of filter coefficients, thus the computation demands \cite{cumming2005digital,richards2010principles}. 

In communication systems, filter banks that consists of several matched filters are needed for the optimum receiver under additive noise \cite{meyr1997digital}. In addition, synchronization is implemented through matched filters. In the synchronization algorithms, relatively long pseudo-random (PN) sequences are often used for low signal to noise ratio (SNR) robustness \cite{pickholtz1982theory,vsajic2011low,li2009low}. In code division multiple access (CDMA) based communication schemes (i.e. GPS), parallel filter banks are widely used as optimum receiver filters \cite{farrell1999global,hofmann2012global}.

Convolutional neural networks also use filter banks for the so called convolutional layers. In these layers, input data are subject to, mostly two dimensional filters \cite{lecun2010convolutional,hinton2012improving}. At each convolutional layer, learned weights form a set of filters. As the number of layers increase, the amount of data processed create a computationally exhausting problem. In addition to this, in neural networks, layers process data in parallel, which leads to parallel filter bank structures. Due to these properties of neural networks, GPUs are often preferred for their high parallel computational power \cite{oh2004gpu,krizhevsky2012imagenet}.

Traditionally, each filter in a parallel filter bank is implemented as a separate filter. This requires increased number of computation resources as the number of filters and number of coefficients per filter increase. Thus, it is important to decrease the computational needs of filter banks, especially on chip applications, where power and area are both limited and shared among many other algorithms. To the best of our knowledge, efficient filter design algorithms focus around efficient implementation of a single filter. Some of these implementation methods are, polyphase filter structures, sharing of numerically similar coefficients, or design of filter coefficients in order to increase the number of shared coefficients or systolic structures that trade increased sampling frequency with decreased resource usage.


Polyphase filters are one such example of efficient filter implementation, if filtering operation is paired with down sampling (decimation) or up sampling (interpolation) \cite{vaidyanathan1990multirate}. This type of implementation methods are widely used in multirate signal processing applications. Decimation operation is used in various digital signal processing applications, from speech processing to digital communications \cite{vaidyanathan1990multirate,harris2003digital,bellanger1976digital}. For example, in conventional interpolation applications, first input data is upsampled then is fed into a filter. This requires filter to process data faster than the original sampling rate of the signal. In polyphase representation upsampling and filtering operations are reversed using Noble's identity. First filtering is applied and then output data is upsampled. This effectively decreases the rate of the filtering operation resulting in a power efficient method. However, polyphase filters simply divide a single filter into parts, effectively implementing them as a parallel filter bank. It does not decrease the number of operations at each clock cycle, thus does not offer area efficiency.

 If the absolute value of filter the coefficients are the same, coefficient sharing within a filter is another approach of optimization \cite{fliege1994multirate,jordan1986correlation,karlsson2005implementation}. Such algorithms are especially useful with symmetric filters. However, this method focuses on optimization of a single filter, not the filter bank as a whole. Moreover, even though the coefficients are shared, they require increased computation rate. These approaches trade number of multiplication operations (i.e. reduce the number of multipliers) with faster computation rate.

Another approach is to optimize the filter coefficients themselves so that they can be shared in an efficient manner. A case of this is the optimization of the filter coefficients itself so that they can be shared using any sharing algorithm. In \cite{parker1997low}, the filter coefficients are quantized in order to retain the frequency properties of the filter and to share the polyphase filter coefficients as much as possible. In \cite{sriram1999low,taylor2014efficient}, a pseudo-random (PN) sequence is specifically designed so that the repeating structures can be efficiently exploited in order to reduce the overall complexity of the filter.   

In filters, the tapped delay line is one of the major resource heavy elements due to number of shift registers it requires. In \cite{li2009low} the filter coefficients are grouped such that a smaller tapped delay line can be shared between groups and thus the length of the tapped delay line is significantly reduced. However, this algorithm also focuses on optimization of a single filter \cite{taylor2014efficient}. In a parallel filter bank, since input is the same for each filter, tapped delay line can be shared among all of the filters, however, this approach also does not decrease the number of computations required.

In this paper, an efficient filter bank design algorithm is presented, which aims to reduce the number of computation operations needed for a set of filters that use PN sequences as filter coefficients. Algorithm presented does not increase the rate of the system like conventional coefficient sharing algorithms as in \cite{fliege1994multirate,jordan1986correlation,karlsson2005implementation}. Additionally, algorithm also does not require exploitation of polyphase representation as in \cite{vaidyanathan1990multirate,harris2003digital,bellanger1976digital}. The algorithm's aim is to share coefficients between a number of filters according to a two step procedure. Coefficients of filters are simply grouped and rearranged. The algorithm significantly reduces the number of addition operations with its novel approach. In addition to this, algorithm enables much more flexible applications. It can be combined with any of the optimization algorithms 	aforementioned in this paper. One major example of coefficient sharing in expense of processing rate is also presented in this paper.

The algorithm can also be extended to filters with any value for their coefficients, from PN sequences by quantizing the absolute value of the coefficients. Such implementation of the algorithm not only reduces the number of summations, but also the number of multiplications.

The paper is organized as follows, in Section \ref{sec:coeff_share}, the algorithm is presented for filters with PN sequences as their coefficients. In Section \ref{sec:optimization_bounds}, optimization bounds of the algorithm is formulated, and in Section \ref{sec:fpga}, FPGA implementation for several cases are shown as an example for the efficiency of the algorithm. Finally we conclude this paper in Section \ref{sec:conclusion}.

\section{Coefficient Sharing between Filters in a Parallel Filter Bank}
\label{sec:coeff_share}
For a PN sequence filter, coefficients are exclusively $\pm 1$, i.e. $h_k[m] = \pm1 \, \forall m \in \mathcal{Z}^+$.

For a one-dimensional signal $x$, filtering is defined through the following convolution operation,
\begin{equation}
	y[n] = x[n] \ast h[n] = \sum_{m=0}^{M} x[n-m]h[m],
	\label{eq:convolution}
\end{equation}
where, $h$ is the filter and $y$ is the output of the filtering operation for this filter. Since, PN filter coefficients are either 1 or -1 we define the following sets:

\begin{subequations}
\begin{equation}
S_{1} = \{m | h[m] = 1\}
\end{equation}
and
\begin{equation}
S_{0} = \{m | h[m] = -1\}
\end{equation}
\label{eq:subsets}
\end{subequations}

Rearranging the summation order in Eq \eqref{eq:convolution}, we obtain the following operation:

\begin{equation}
	y[n] = \sum_{m\in S_1} x[n-m] - \sum_{m\in S_0} x[n-m]
	\label{eq:convolutionExtended}
\end{equation}

The total number of summations to implement Eq \eqref{eq:convolutionExtended} is equal to the summations in Eq \eqref{eq:convolution}. Rearrangement of the order of the summations as in Eq \eqref{eq:convolutionExtended} would only require proper handling of the tapped delay line of the filter structure which does not introduce any additional resource requirement.

Let's assume there are two parallel filters in a filter bank: $h_1$ and $h_2$. From the coefficients of such filters we define the following four sets:

\begin{subequations}
\begin{equation}
S_{00} = \{m | h_1[m] = -1\ \&\ h_2[m] = -1\},
\end{equation}
\begin{equation}
S_{01} = \{m | h_1[m] = 1\ \&\ h_2[m] = -1\},
\end{equation}
\begin{equation}
S_{10} = \{m | h_1[m] = -1\ \&\ h_2[m] = 1\},
\end{equation}
\begin{equation}
S_{11} = \{m | h_1[m] = 1\ \&\ h_2[m] = 1\}.
\end{equation}
\label{eq:subsetss}
\end{subequations}

We then rewrite the convolution operations for these two filters as follows:

\begin{subequations}
\begin{multline}
y_1[n] = -\sum_{m\in S_{00}} x[n-m] + \sum_{m\in S_{01}} x[n-m] \\
-\sum_{m\in S_{10}} x[n-m] + \sum_{m\in S_{11}} x[n-m]
\end{multline}
and
\begin{multline}
	y_2[n] =  -\sum_{m\in S_{00}} x[n-m] -\sum_{m\in S_{01}} x[n-m] \\
	+ \sum_{m\in S_{10}} x[n-m] + \sum_{m\in S_{11}} x[n-m]
\end{multline}
\label{eq:convolutionTwoFilters}
\end{subequations}

From Eq \eqref{eq:convolutionTwoFilters}, we see that in order to implement two different filters in a parallel filter bank we needed to implement only four distinct summations. We then sum or subtract the results of these summations in order to obtain $y_1$ and $y_2$. We extend the idea further into $K$ filters. For this we first construct the indices set for each filter as follows:
\begin{subequations}
\begin{equation}
	S_0^{k} = \{m | h_k[m] = -1\}
\end{equation}
and
\begin{equation}
	S_1^{k} = \{m | h_k[m] = 1\},
\end{equation}
\end{subequations}
where $k$ is the index of the filter. Using these sets, we define the following intersection sets similar to the ones in Eq \eqref{eq:subsetss}:
\small
\begin{equation}
\begin{bmatrix}
S_{00..00}\\S_{00..01}\\S_{00..10}\\ \vdots \\ S_{11..11}
\end{bmatrix} = 
\begin{bmatrix*}
S_0^{1} \cap S_0^{2}  \cap \hdots \cap S_0^{K-1}  \cap S_0^{K}  \\
S_1^{1} \cap S_0^{2}  \cap \hdots \cap S_0^{K-1}  \cap S_0^{K}  \\
S_0^{1} \cap S_1^{2}  \cap \hdots \cap S_0^{K-1}  \cap S_0^{K}  \\
\vdots \\
S_1^{1} \cap S_1^{2}  \cap \hdots \cap S_1^{K-1}  \cap S_1^{K}  \\
\end{bmatrix*}
\label{eq:intersection}
\end{equation} 
\normalsize

The number of subsets is approximately equal to $2^K$. This number can grow fast as the number of parallel filters increase. The efficiency of the proposed algorithm compared to the number of filters and filter size is discussed Section \ref{sec:optimization_bounds}.

Eq \eqref{eq:filter_bank_example} shows the summations over the defined subsets in order to calculate the filter results for each individual filter.

\begin{figure*}[!t]
\begin{equation}
\begin{bmatrix}
y_1[n]\\y_2[n]\\ \vdots \\ y_{K}[n]
\end{bmatrix} = 
\begin{bmatrix*}
-\sum\limits_{m\in S_{00..00}} x[n-m]+\sum\limits_{m\in S_{00..01}} x[n-m]+ \dots  -\sum\limits_{m\in S_{11..10}} x[n-m]+\sum\limits_{m\in S_{11..11}} x[n-m]\\
-\sum\limits_{m\in S_{00..00}} x[n-m]-\sum\limits_{m\in S_{00..01}} x[n-m]+ \dots  +\sum\limits_{m\in S_{11..10}} x[n-m]+\sum\limits_{m\in S_{11..11}} x[n-m]\\

\vdots \\
-\sum\limits_{m\in S_{00..00}} x[n-m]-\sum\limits_{m\in S_{00..01}} x[n-m]+ \dots  +\sum\limits_{m\in S_{11..10}} x[n-m]+\sum\limits_{m\in S_{11..11}} x[n-m]
\end{bmatrix*}
\label{eq:filter_bank_example}
\end{equation} 
\normalsize
\hrulefill
\end{figure*}

Calculation of summations in one row of the Eq \eqref{eq:filter_bank_example} is sufficient to calculate results of each filter since these sums repeat themselves in each row with a different sign. The sets defined in Eq \eqref{eq:intersection} are exclusive. Therefore the total number of summation operations defined in a row of Eq \eqref{eq:filter_bank_example} is equal to the number of coefficients of filters in the filter bank. In Eq \eqref{eq:filter_bank_example}, the filtering operation is divided into two main structures. In the first summation structure, outputs of subset summations, as found in Eq \eqref{eq:intersection}, are calculated. In the second structure, subsets are summed once more to calculate the actual filter outputs.

At this stage we reduce the problem into summations of $2^K$ values for each filter. When the number of filters is significantly smaller than the number of coefficients in the filters, Eq \eqref{eq:filter_bank_example} reduces the total number of summations required to implement the filter bank.

\section{Optimization Bounds of Coefficient Sharing in Filter Bank}
\label{sec:optimization_bounds}
In this section, the optimization performance of the algorithm is discussed. The main focus of efficiency in the proposed method is the reduction of the number of operations needed.

There are two ways to implement Eq \eqref{eq:filter_bank_example}, the outer summations can be either summed using a simple summation pyramid as in \cite{hawkes2005dsp} or using multiply-and-accumulate (MAC) structure to map the operations to DSP48 blocks. Optimization bounds for both of the methods are presented in this section.

In order to give a reasonable bound for the optimization performance of the algorithm, all the filter coefficients of the filter bank are assumed to be taken from independent and identically distributed (iid) Bernoulli trials with equal probability of $1$ and $-1$ outcomes.

First assume that there are $K$ filters and $M$ coefficients in each filter with coefficients $\pm 1$. As shown previously in Section \ref{sec:coeff_share}, a filter bank with $K$ filters will have at most $2^{K}$ coefficient subsets. Statistically, these coefficient subsets will all have equal probability of occurring since coefficients are all iid. Number of coefficient subsets are important since it directly affects the number of operations needed where the subsets are connected to form the actual filter outputs as given in Eq \eqref{eq:filter_bank_example}.

It is desirable to have each coefficient subset to have high number of elements, since this means more coefficients can be shared. This is possible by either keeping $K$ small or $M$ high. Average number of elements of each coefficient subset can be written as follows:
\begin{equation}
    E[|S_{00..00}| + ... + |S_{11..11}|] = \frac{M}{2^K},
    \label{eq:average_elements}
\end{equation}
where, $2^K$ is the total number of subsets. For a given $M$, large $K$ results in better optimization considering the $M$ also large. However as $K$ grows, the number of sets in \eqref{eq:intersection} grows exponentially making the algorithm impractical. It is evident that there is convex optimization surface and it is desirable to find a proper relationship between $K$ and $M$.
 
\subsection{Number of Operations for the Filter Bank}
For an $M/2^K$ number of coefficients in a subset, we need $M/2^K$ MAC operations. Then the expected number of MACs needed to implement all the subsets in a filter bank is given Eq \eqref{eq:num_mul_acc_1}.

\begin{equation}
E\big[\Pi_{1}\big] = \bigg(\frac{M}{2^{K}} \bigg)\times 2^{K} = M 
\label{eq:num_mul_acc_1}
\end{equation}

With this many MACs, all the subsets are implemented, however additional operations are needed to implement the outer summations given in Eq \eqref{eq:filter_bank_example}. Outer summations can be computed using either MAC blocks or a summation pyramid as in \cite{hawkes2005dsp}. Both approaches have their own advantages. Summation pyramid naturally does not use extra MAC operations and thus uses less DSP48s. However, multiply-and-accumulate implementation enables systolic implementation at the cost of increased sampling frequency.

\subsubsection{Implementation of Outer Summations using MACs}
Total number of MACs needed to implement outer summations in the filter bank is as follows:
\begin{equation}
E\big[\Pi_{2} \big] = (2^{K}) \times K
\label{eq:num_mul_acc_2}  
\end{equation}
where, first term of the multiplication is the number of MAC operations needed to implement a single filter in the filter bank and second term of the multiplication is the total number of filters in the filter bank.

Hence, the expected number of MAC operations needed is in Eq \eqref{eq:total_mul_acc_1}.
\begin{equation}
\begin{split}
E\big[\Pi \big] & = E\big[\Pi_{1} \big]  + E\big[\Pi_{2} \big] \\
                   & = M + (2^{K}) \times K,
\end{split}
\label{eq:total_mul_acc_1}
\end{equation}
where the optimization can be done if either of $K$ or $M$ is not given. 
\begin{equation}
\begin{aligned}
\min_{K,M} \quad & E\big[\Pi \big], \\
\textrm{s.t.} \quad & \frac{M}{2^{K}} \gg 1\\
\end{aligned}
\label{eq:minimize_mul_acc_1}
\end{equation}

\subsubsection{Implementation of Outer Summations using Summation Pyramids}
Since inner summations are simply summed or subtracted, it is also possible to not use multiply-and-accumulate operations at all. In order to achieve this, a summation pyramid as in \cite{hawkes2005dsp} can be built that uses two input summation operations. 

Total number of two input summations needed to implement outer summations in the filter bank is as follows:
\begin{equation}
E\big[\Sigma_{2} \big] = (2^{K}-1) \times K
\label{eq:num_sum_2}  
\end{equation}
where, first term of the multiplication is the number of two input summation operations needed to implement a single filter in the filter bank and second term of the multiplication is the total number of filters in the filter bank.
Hence, the expected number of operations needed is in Eq \eqref{eq:total_sum_1}.
\begin{equation}
\begin{split}
E\big[O \big] & = E\big[\Pi_{1} \big]  + E\big[\Sigma_{2} \big] \\
                   & = M + (2^{K}-1) \times K,
\end{split}
\label{eq:total_sum_1}
\end{equation}
where $E\big[\Pi_{1} \big]$ term comes from the number of MAC operations needed to compute the subsets. The optimization can be done if either of $K$ or $M$ is not given. 
\begin{equation}
\begin{aligned}
\min_{K,M} \quad & E\big[O \big], \\
\textrm{s.t.} \quad & \frac{M}{2^{K}} \gg 1\\
\end{aligned}
\label{eq:minimize_sum_1}
\end{equation}
An important note is that Eq \eqref{eq:total_sum_1} is the total number, multiply-and-accumulate and two input summations, hence an analysis of the number of each operation is valuable.

In most practical cases, $K$ and $M$ both are pre-defined according to the needs of the application, thus Eq \eqref{eq:total_mul_acc_1} and Eq \eqref{eq:total_sum_1} are simply the number of total operations needed for such a case and no optimization can be done. Another problem of the algorithm is that, when $\frac{M}{2^{K}}$ is comparable to $1$, the total number of operations may be higher than the operations needed in a regular filter bank.

In order to overcome these problems and efficiently optimize the filters, another term, $G$ is defined as the number of groups in which filters in the filter bank are grouped together to create sub-filter banks. Grouping filters in any combination does not have an additional impact on the algorithm due to the iid coefficients assumption.

\subsection{Grouping Filters into Smaller Filter Banks}
A filter bank with $K$ filters grouped into $G$ groups means there are $G$ filter banks with $\frac{K}{G}$ filters each. Each filter bank group then have $2^{(K/G)}$ coefficient subsets as defined previously in Eq \eqref{eq:intersection}. Statistically, each subset will have $\frac{M}{2^{(K/G)}}$ elements that need to be summed. 

Expected number of MAC operations needed for the first stage of the algorithm for a group of the filter bank is in Eq \eqref{eq:pi_1}.
\begin{equation}
    E\big[\Pi_{1}(G)\big] = \bigg(\frac{M}{2^{(K/G)}}\bigg)\times 2^{(K/G)} = M, 
    \label{eq:pi_1}
\end{equation}

\subsubsection{Implementation of Outer Summations using Multiply-and-Accumulates}
The rule defined in Eq \eqref{eq:convolutionTwoFilters} can be implemented using MAC operations. The expected number of operations needed is in Eq \eqref{eq:pi_2}.
\begin{equation}
    E\big[\Pi_{2}(G) \big] = (2^{(K/G)}) \times \frac{K}{G}
    \label{eq:pi_2}
\end{equation}
First term of the multiplication in Eq \eqref{eq:pi_2} is the number of multiply-and-accumulate operations needed to implement a single filter in the filter bank group and second term of the multiplication is the total number of filters in the filter bank group.

Total number of multiply-and-accumulate operations needed to implement the whole filter bank is
\begin{equation}
\begin{split}
E\big[\Pi(G) \big] & = \bigg(E\big[\Pi_{1}(G)\big] + E\big[\Pi_{2}(G)\big] \bigg)\times G \\
                   & = G\times M + (2^{(K/G)}) \times K, 
\end{split}
\label{eq:total_mul_acc}
\end{equation}
where $K$ is the total number of filters in the filter bank, $M$ is  the number of coefficients in each filter, $G$ is the number of filter bank groups with each filter bank group containing $\frac{K}{G}$ filters.

Eq \eqref{eq:total_mul_acc} is a convex function. Then, the optimization problem turns into the minimization in Eq \eqref{eq:minimize_mul_acc}.
\begin{equation}
\begin{aligned}
\min_{G} \quad & E\big[\Pi(G) \big], \\
\textrm{s.t.} \quad & \frac{M}{2^{K/G}} \gg 1\\
\end{aligned}
\label{eq:minimize_mul_acc}
\end{equation}

\subsubsection{Implementation of Outer Summations using Summation Pyramids}
In order to implement the filters, the output of the subsets are summed according to the rule defined in Eq \eqref{eq:convolutionTwoFilters} and the filter coefficients are shared among filters of the filter bank group. For this stage, the expected number of summation operations needed is in Eq \eqref{eq:sigma_2}.
\begin{equation}
    E\big[\Sigma_{2}(G) \big] = (2^{(K/G)}-1) \times \frac{K}{G}
    \label{eq:sigma_2}
\end{equation}
First term of the multiplication in Eq \eqref{eq:sigma_2} is the number of two input summation operations needed to implement a single filter in the filter bank group and second term of the multiplication is the total number of filters in the filter bank group.

Finally, total number of operations needed to implement the whole filter bank is
\begin{equation}
\begin{split}
E\big[O(G) \big] & = \bigg(E\big[\Pi_{1}(G)\big] + E\big[\Sigma_{2}(G)\big] \bigg)\times G \\
                   & = G\times M + K\times(2^{(K/G)}-1), 
\end{split}
\label{eq:total_sum}
\end{equation}
where $K$ is the total number of filters in the filter bank, $M$ is  the number of coefficients in each filter, $G$ is the number of filter bank groups with each filter bank group containing $\frac{K}{G}$ filters. 

The optimization of the algorithm is done through minimizing the expected number of summation operations. Eq \eqref{eq:total_sum} is a convex function. Then, the optimization problem turns into the minimization in Eq \eqref{eq:minimize}.
\begin{equation}
\begin{aligned}
\min_{G} \quad & E\big[O(G) \big], \\
\textrm{s.t.} \quad & \frac{M}{2^{K/G}} \gg 1\\
\end{aligned}
\label{eq:minimize}
\end{equation}
An important note is that, $O(G)$ is the total number of operations, multiply-and-accumulates and two input summations combined. Hence an analysis of individual operations is also noteworthy and will be presented.

\begin{figure}[h]     
	\centering
	\includegraphics[width=0.48\textwidth]{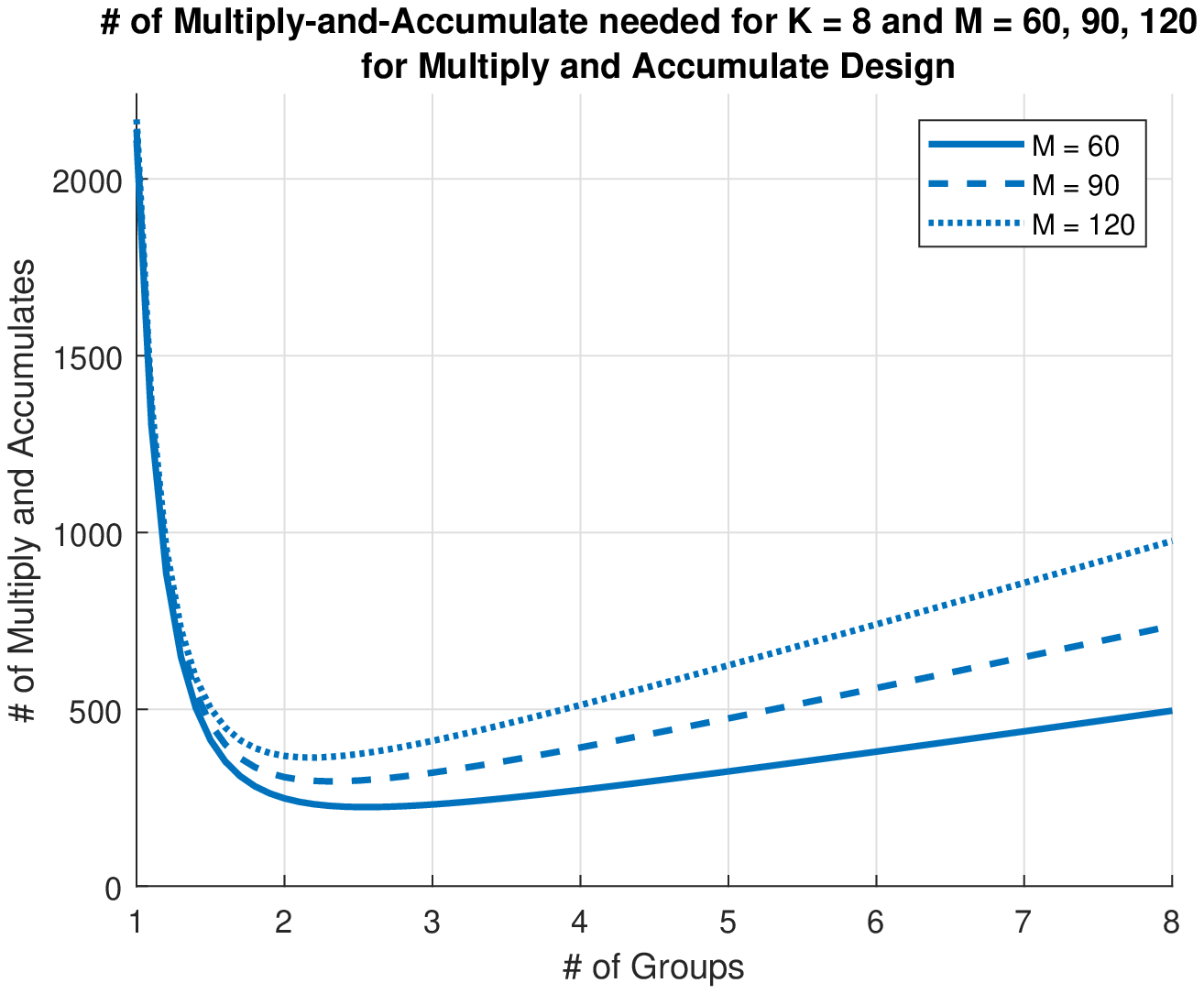}
	\caption{\# of MAC operations required.}
	\caption{Optimization of Eq \eqref{eq:minimize_mul_acc} with respect to $G$ for various $M$ and $K=8$.}
	\label{fig:minimize_curve_mul_acc}
\end{figure}

\begin{figure}[h]     
	\begin{subfigure}[h]{0.48\textwidth}
	\centering
	\includegraphics[width=1\textwidth]{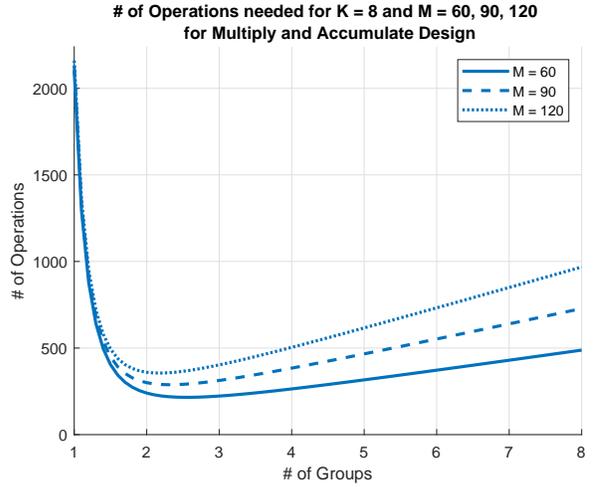}
	\caption{Total \# of operations required.}
	\label{fig:total_op_1}
	\end{subfigure}
	
	\begin{subfigure}[h]{0.48\textwidth}
	\centering
	\includegraphics[width=1\textwidth]{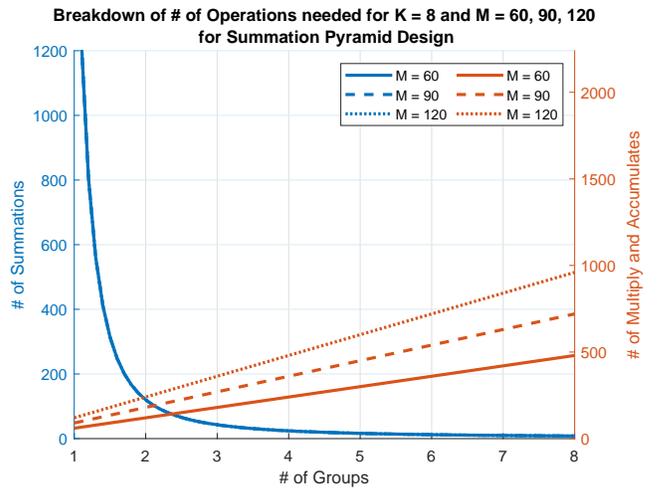}
	\caption{Analysis of the number of operations.}
	\end{subfigure}
	\caption{Optimization of Eq \eqref{eq:minimize} with respect to $G$ for various $M$ and $K=8$.}
	\label{fig:minimize_curve}
\end{figure}

The solution of the minimization in Eq \eqref{eq:minimize_mul_acc} and Eq \eqref{eq:minimize} for some example cases are in Fig. \ref{fig:minimize_curve_mul_acc} and Fig. \ref{fig:minimize_curve}, respectively. From Fig. \ref{fig:minimize_curve}, it can be quickly realized that decreasing $2^{K/G}$ as much as possible does not decrease the total number of operations necessarily. This is because the algorithm has two stages and decreasing the number of operations in first stage increases the operations in second stage and vice verse. Fig. \ref{fig:minimize_curve_mul_acc} and Fig. \ref{fig:total_op_1} are essentially same, because in Fig. \ref{fig:minimize_curve_mul_acc}, two input summations are simply converted to MAC operations.

This optimization is a crude one since it assumes there can be fractional number of filters in a group of filter bank. Thus, a realistic optimization bound is also needed.

\subsection{Discrete Grouping of Filters}
In a realistic case, $\frac{K}{G}$ cannot be a fractional number. The optimization problem is modified using a discrete approach as follows:
\begin{enumerate}
    \item $G_1 = mod(K, G)$ amount of filter bank groups will have $|G_1| = \frac{K-mod(K,G)}{G}+1$ filters. For these filter bank groups, the expected number of MAC operations in the first stage is in Eq \eqref{eq:g_1_sigma_1} for each filter bank group,
    \begin{equation}
    \begin{split}
    E\big[\Pi_{1}(G_1)\big] & = \bigg(\frac{M}{2^{|G_1|}}\bigg) 2^{|G_1|} \\ 
                               & = M 
    \end{split}
    \label{eq:g_1_sigma_1}
    \end{equation}
    \normalsize
    
    \textbullet \, If the filter stage is implemented using MAC blocks, expected number of operations in the second stage for the first $G_1$ filter bank groups is in Eq \eqref{eq:g_1_pi_2} for each filter bank group.
    \begin{equation}
        E\big[\Pi_{2}(G_1) \big] = (2^{|G_1|}) |G_1|
        \label{eq:g_1_pi_2}
    \end{equation}
    and the total number of MACs will be,
    \begin{equation}
    E\big[\Pi(G_1) \big] = \bigg(E\big[\Pi_{1}(G_1)\big] + E\big[\Pi_{2}(G_1)\big] \bigg)  G_1,
    \label{eq:total_pi_G_1} 
    \end{equation}
    
    \textbullet \, If the filter stage is implemented using summation blocks, expected number of summation operations in the second stage for the first $G_1$ filter bank groups is in Eq \eqref{eq:g_1_sigma_2} for each filter bank group.
    \begin{equation}
        E\big[\Sigma_{2}(G_1) \big] = (2^{|G_1|}-1) |G_1|
        \label{eq:g_1_sigma_2}
    \end{equation}
    and the expected number of operations will be,
    \begin{equation}
    E\big[O(G_1) \big] = \bigg(E\big[\Pi_{1}(G_1)\big] + E\big[\Sigma_{2}(G_1)\big] \bigg)  (G_1),
    \label{eq:total_sum_G_1}
    \end{equation}
    \normalsize
    Eq \eqref{eq:total_sum_G_1} is the total number of operations.
    
    \item Rest of the $G_2 = G-mod(K,G)$ amount of filter bank groups will have $|G_2| = \frac{K-mod(K,G)}{G}$ amount of filters. Then the first stage will have expected number of MACs as in Eq \eqref{eq:g_2_pi_1},
    \begin{equation}
    \begin{split}
    E\big[\Pi_{1}(G_2)\big] & = \bigg(\frac{M}{2^{|G_2|}}\bigg) 2^{|G_2|} \\ 
                               & = M
    \end{split}
    \label{eq:g_2_pi_1}
    \end{equation}
    
    \textbullet \, If the filter stage is implemented using MAC blocks, expected number of operations in the second stage for this filter bank group is in Eq \eqref{eq:g_2_pi_2} for each filter bank group.
    \begin{equation}
    E\big[\Sigma_{2}(G_2) \big] = (2^{|G_2|}) |G_2|
    \label{eq:g_2_pi_2}
    \end{equation}
    and total number of MAC operations will be,
    \begin{equation}
    E\big[\Pi(G_2) \big] = \bigg(E\big[\Pi_{1}(G_2)\big] + E\big[\Pi_{2}(G_2)\big] \bigg)  (G_2).
    \label{eq:total_pi_G_2}
    \end{equation}
    
    \textbullet \, If the filter stage is implemented using summation blocks, expected number of summation operations in the second stage for this filter bank group is in Eq \eqref{eq:g_2_sigma_2} for each filter bank group.
    \begin{equation}
    E\big[\Sigma_{2}(G_2) \big] = (2^{|G_2|}-1) |G_2|
    \label{eq:g_2_sigma_2}
    \end{equation}
    and total number of operations will be,
    \begin{equation}
    E\big[O(G_2) \big] = \bigg(E\big[\Pi_{1}(G_2)\big] + E\big[\Sigma_{2}(G_2)\big] \bigg)  (G_2).
    \label{eq:total_sum_G_2}
    \end{equation}
    Again, Eq \eqref{eq:total_sum_G_2} is the total number of operations.
\end{enumerate}

For the full MAC implementation, using Eq \eqref{eq:total_pi_G_1} and \eqref{eq:total_pi_G_2} the total number of multiply-and-accumulates needed to implement the filter bank can be found as follows:
\begin{equation}
E\big[\Pi \big] = E\big[\Pi(G_1) \big]  + E\big[\Pi(G_2) \big].
\end{equation}
The optimization problem is then,
\begin{equation}
\begin{aligned}
\min_{G}      \quad & E\big[\Pi(G_1) \big]  + E\big[\Pi(G_1) \big], \\
\textrm{s.t.} \quad & \frac{M}{2^{|G_1|}} \gg 1\\
                    & \frac{M}{2^{|G_2|}} \gg 1 \\
\end{aligned}
\label{eq:minimize_real_pi}
\end{equation}
where $|G_1| = \frac{K-mod(K,G)}{G}+1$ and $|G_2| = \frac{K-mod(K,G)}{G}$.

For the hybrid MAC and summation pyramid implementation, \eqref{eq:total_sum_G_1} and Eq \eqref{eq:total_sum_G_2}, the total number of summations needed to implement this filter bank can be found as follows:
\begin{equation}
E\big[O \big] = E\big[O(G_1) \big]  + E\big[O(G_2) \big].
\end{equation}
The optimization problem is then,
\begin{equation}
\begin{aligned}
\min_{G}      \quad & E\big[O(G_1) \big]  + E\big[O(G_1) \big], \\
\textrm{s.t.} \quad & \frac{M}{2^{|G_1|}} \gg 1\\
                    & \frac{M}{2^{|G_2|}} \gg 1 \\
\end{aligned}
\label{eq:minimize_real}
\end{equation}
where $|G_1| = \frac{K-mod(K,G)}{G}+1$ and $|G_2| = \frac{K-mod(K,G)}{G}$.
Eq \eqref{eq:minimize_real} is the total number of operations and a breakdown of individual operations is valuable.

\begin{figure}[h]     
	\centering
	\includegraphics[width=0.48\textwidth]{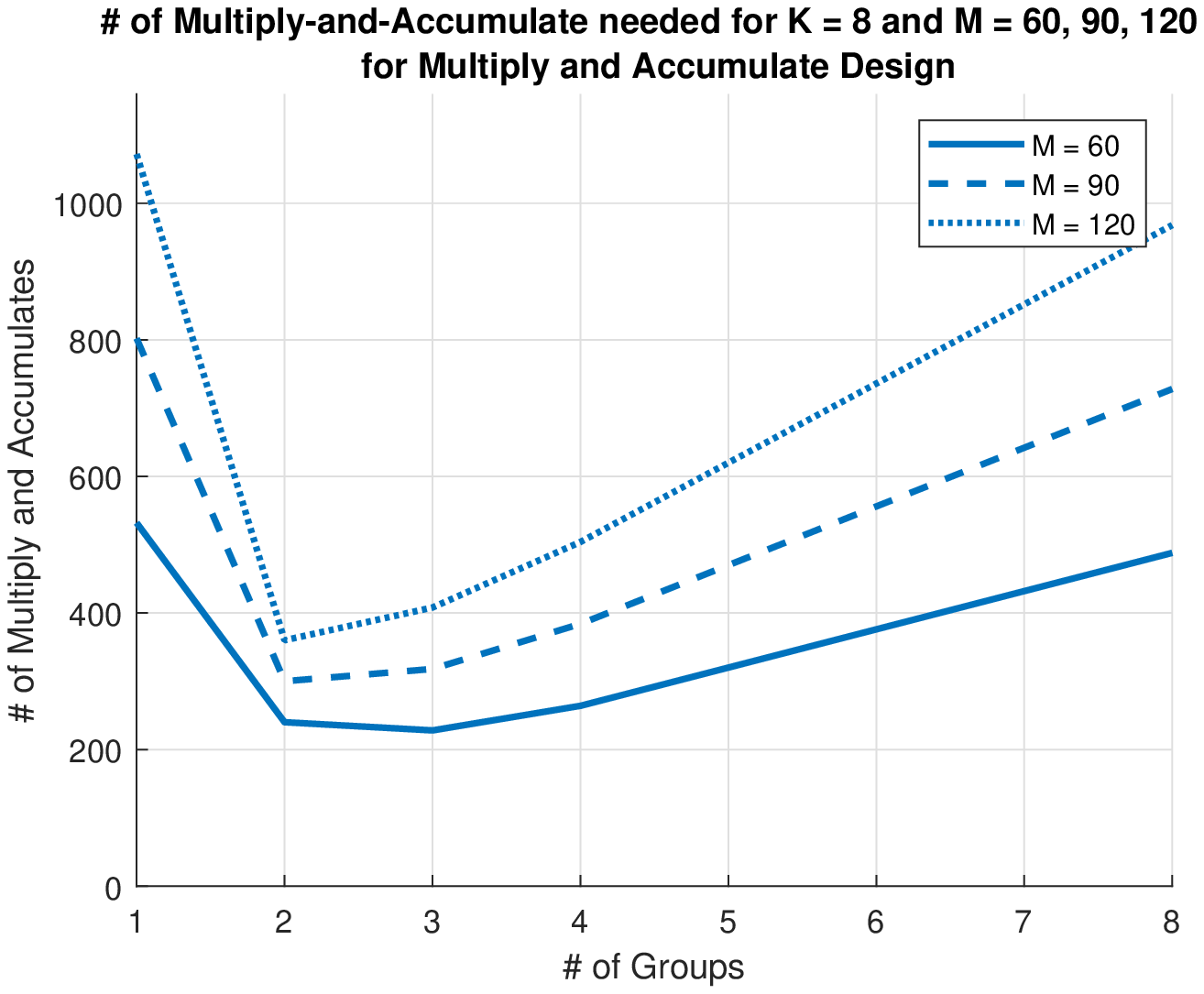}
	\caption{\# of MAC operations required.}
	\caption{Optimization of Eq \eqref{eq:minimize_real_pi} with respect to $G$ for various $M$ and $K=8$.}
	\label{fig:minimize_curve_mul_acc_real}
\end{figure}

\begin{figure}[h]     
	\begin{subfigure}[h]{0.48\textwidth}
	\centering
	\includegraphics[width=1\textwidth]{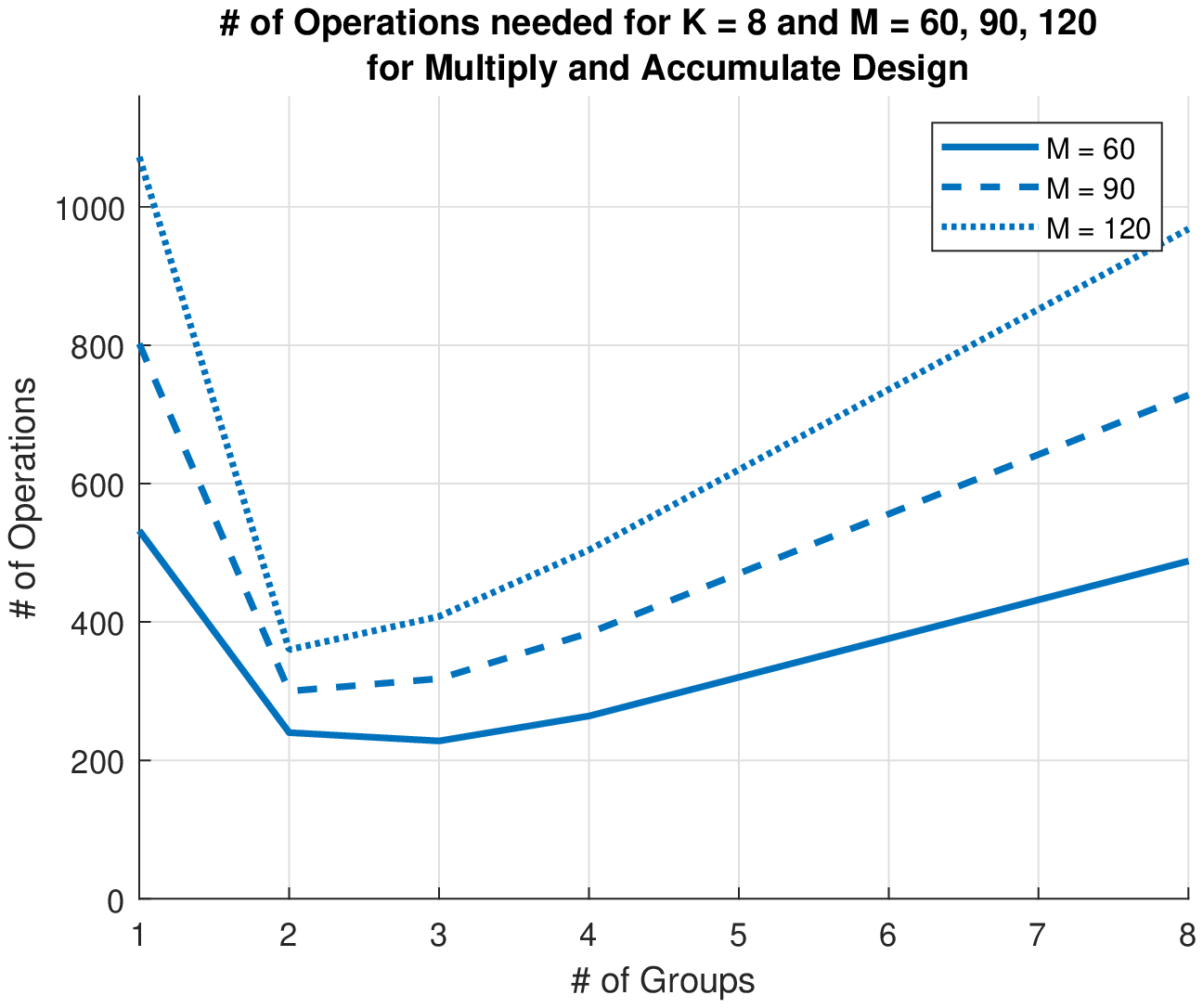}
	\caption{Total \# of operations required.}
	\label{fig:total_op_real_1}
	\end{subfigure}
	
	\begin{subfigure}[h]{0.48\textwidth}
	\centering
	\includegraphics[width=1\textwidth]{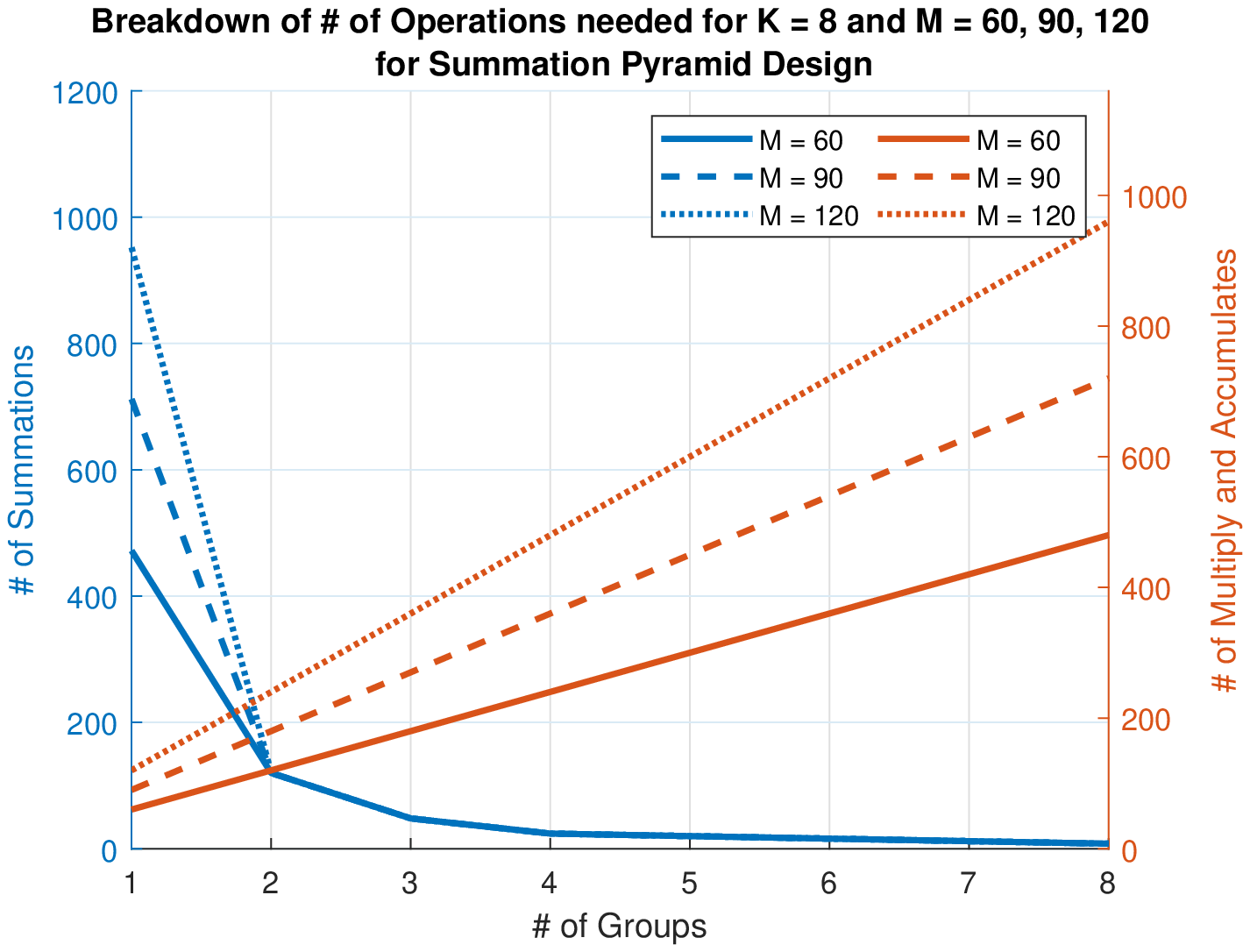}
	\caption{Analysis of the number of operations.}
	\end{subfigure}
	\caption{Optimization of Eq \eqref{eq:minimize_real} with respect to $G$ for various $M$ and $K=8$.}
	\label{fig:minimize_curve_real}
\end{figure}

An example of Eq \eqref{eq:minimize_real_pi} and \eqref{eq:minimize_real} is in Fig. \ref{fig:minimize_curve_mul_acc_real} and \ref{fig:minimize_curve_real}. The optimization problem is again convex and have clear minimums.

\section{FPGA Implementation Preliminaries}
\label{sec:fpga}
In this section, FPGA synthesis performance of the algorithm is compared to other filter design algorithms referenced in Section \ref{sec:intro}. Comparison is grouped under two main applications, a) FIR filter bank, b) Polyphase filter bank. In each application, other coefficient sharing methods referenced in Sec. \ref{sec:intro} are also implemented. In order to achieve fast prototyping and considering the complexity of our algorithm to implement on FPGA, MATLAB's HDL Coder Toolbox is used to generate the necessary HDL codes. All of the coefficient sharing algorithms are also implemented in MATLAB for a fair comparison. Generated HDL codes are then synthesized in Vivado in order to find the resource cost on an FPGA. 

In all sections, direct form FIR design, partially serial systolic architecture, fully serial systolic architecture and our coefficient sharing algorithm are compared with each other. Comparison criteria are, number of lookup tables (LUTs), registers, DSP48s, samping frequency ($F_s$) of the fastest element in the block and finally the output delay of the algorithm with respect to the Direct form FIR implementation. An algorithm is considered to be efficient if it has low values in all of these criteria.

In the designs for all of the different algorithms, tapped delay line is shared among all filters, in order to further reduce resource cost. In addition to this, for partially systolic architecture, upsample blocks are also shared.

\begin{figure}[h]     
	\begin{subfigure}[h]{0.48\textwidth}
    	\centering
    	\includegraphics[width=1\textwidth]{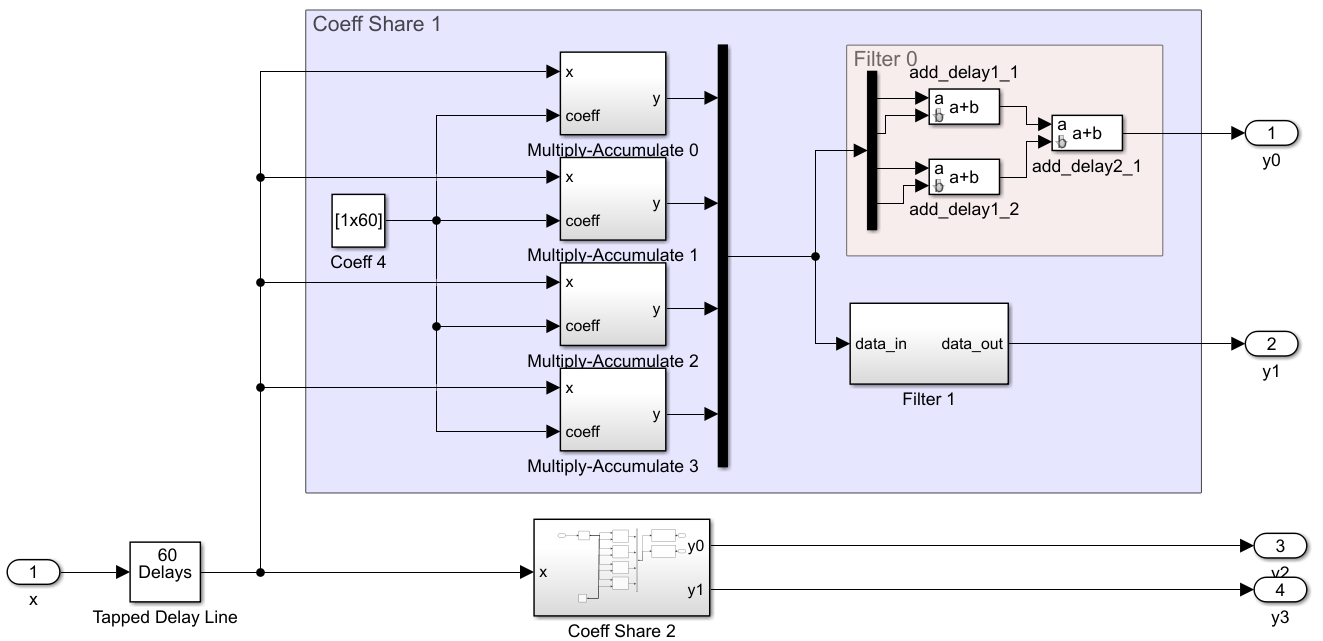}
		\caption{Proposed coefficient sharing algorithm with summation pyramid at \textit{Filter} stage.}
		\label{fig:coeff_share}
	\end{subfigure}
	
	\begin{subfigure}[h]{0.48\textwidth}
    	\centering
    	\includegraphics[width=1\textwidth]{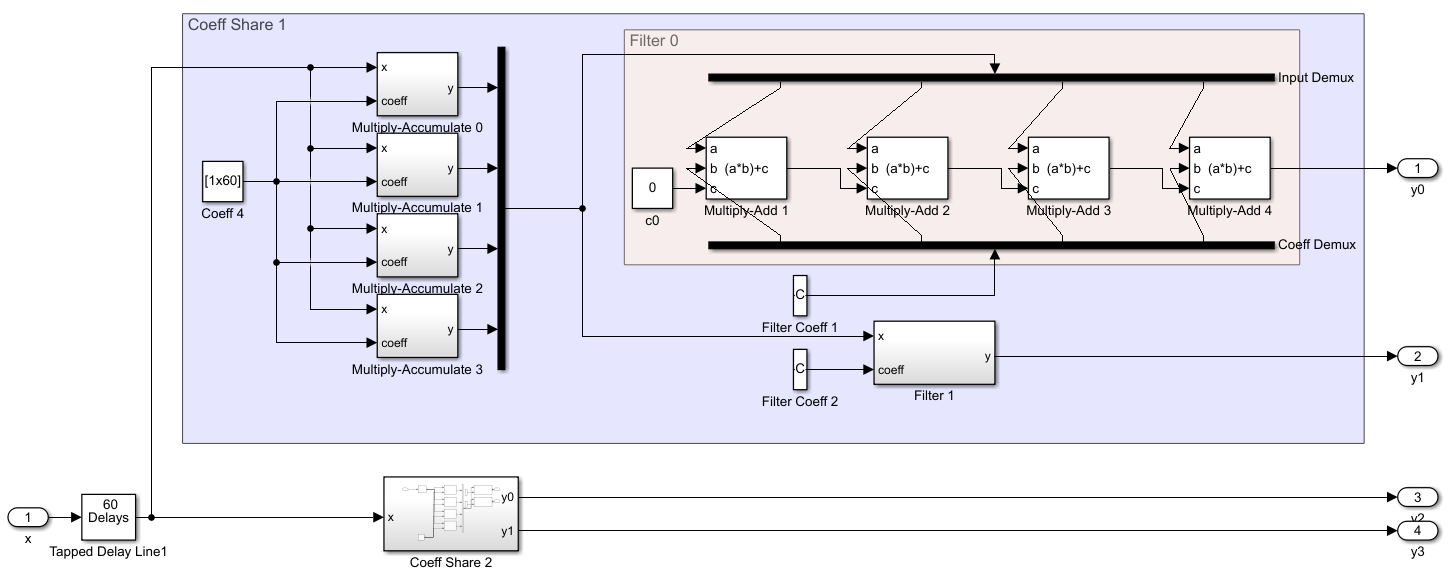}
		\caption{Proposed coefficient sharing algorithm with MAC structure at \textit{Filter} stage}
		\label{fig:coeff_share_dsp48}
	\end{subfigure}
	
	\begin{subfigure}[h]{0.48\textwidth}
    	\centering
    	\includegraphics[width=1\textwidth]{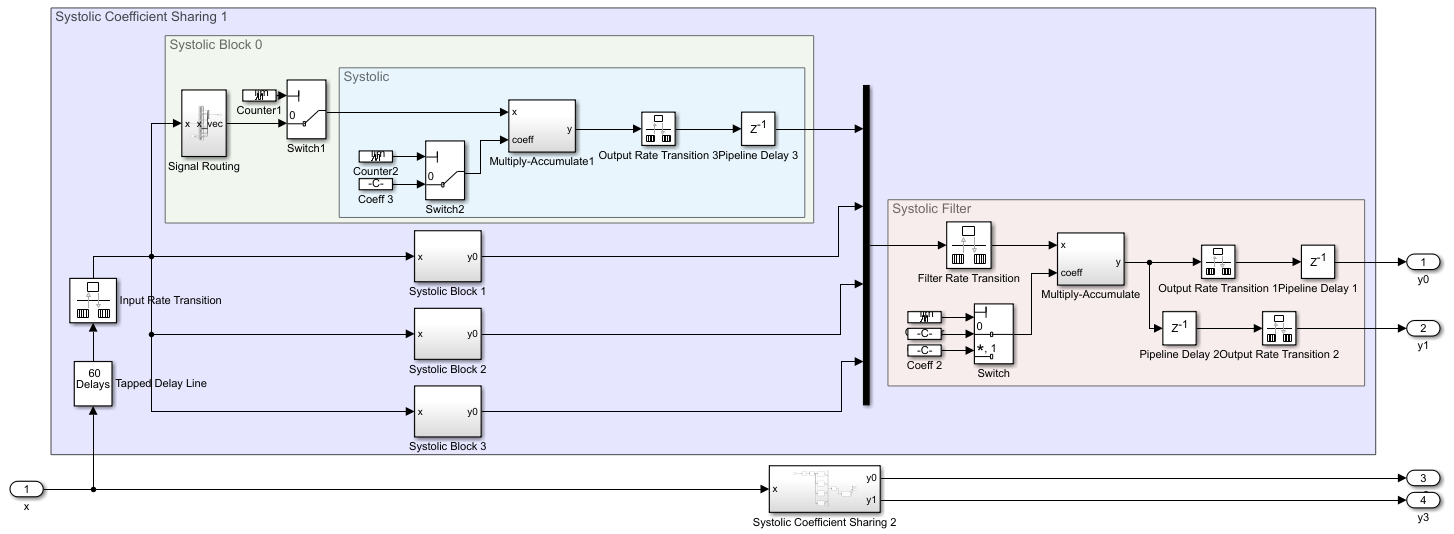}
		\caption{Proposed coefficient sharing algorithm with MAC structure implemented via systolic approach.}
		\label{fig:systolic_coeff_share}
	\end{subfigure}
	
	\caption{MATLAB Simulink block diagrams of proposed coefficient sharing algorithms.}
	\label{fig:implemented_designs1}
\end{figure}

\begin{figure}[h]   
	\begin{subfigure}[h]{0.48\textwidth}
    	\centering
    	\includegraphics[width=1\textwidth]{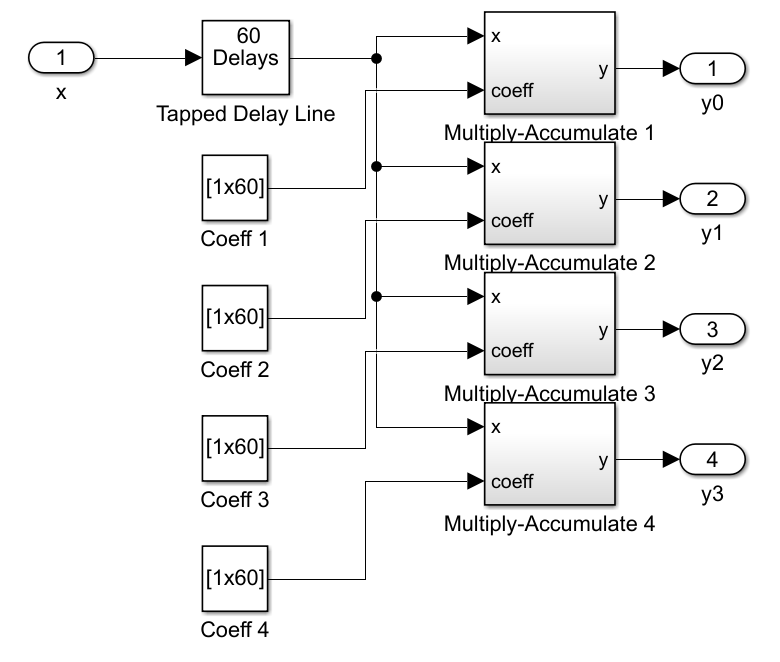}
		\caption{Direct FIR filter with shared tapped delay line.}
		\label{fig:direct_fir}
	\end{subfigure}

	\begin{subfigure}[h]{0.48\textwidth}
    	\centering
    	\includegraphics[width=1\textwidth]{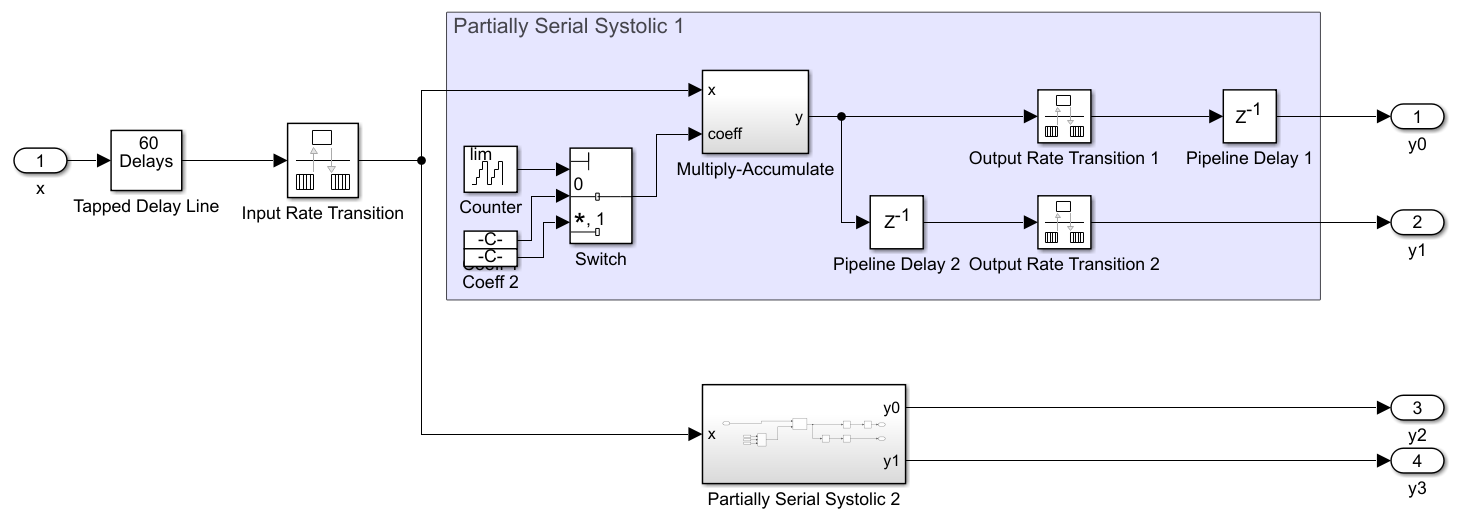}
		\caption{Partially serial systolic structure.}
		\label{fig:partially_serial_systolic}
	\end{subfigure}
	
	\begin{subfigure}[h]{0.48\textwidth}
    	\centering
    	\includegraphics[width=1\textwidth]{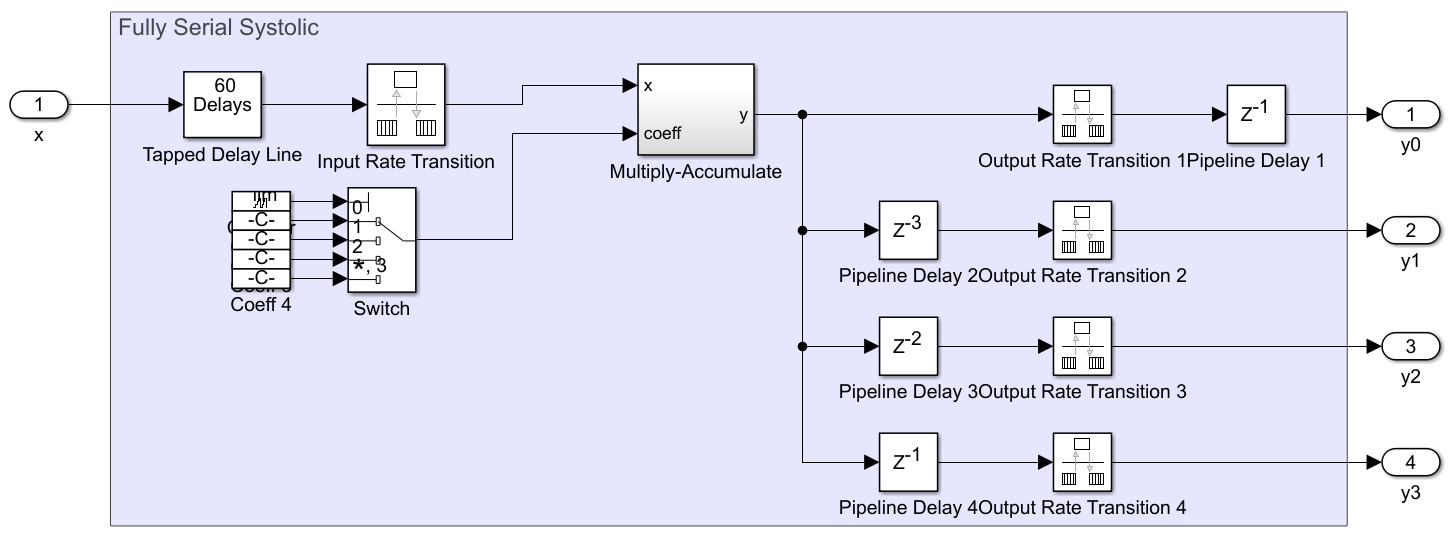}
		\caption{Fully serial systolic structure.}
		\label{fig:fully_serial_systolic}
	\end{subfigure}
	
	\caption{MATLAB Simulink block diagrams of direct form FIR filter and different systolic architectures.}
	\label{fig:implemented_designs2}
\end{figure}

For a sample filter bank with $K = 4$ and $M = 60$, block diagram examples of implemented designs in MATLAB Simulink are in Fig. \ref{fig:coeff_share} \ref{fig:coeff_share_dsp48}, \ref{fig:systolic_coeff_share}, \ref{fig:direct_fir}, \ref{fig:partially_serial_systolic} and \ref{fig:fully_serial_systolic}. In \ref{fig:coeff_share} and \ref{fig:coeff_share_dsp48}, difference is in the \textit{Filter} blocks. In \ref{fig:coeff_share} they are implemented via summation tree as depicted in \cite{hawkes2005dsp} and in \ref{fig:coeff_share_dsp48} they are implemented via MAC structure to map them to DSP48 blocks on FPGA. In Fig. \ref{fig:systolic_coeff_share}, MAC structures are implemented via systolic approach in order to show the flexibility of the proposed algorithm.

In this example block diagrams, number of subgroups for Figs. \ref{fig:coeff_share}, \ref{fig:coeff_share_dsp48} and \ref{fig:systolic_coeff_share} are found via the optimization method presented in Sec. \ref{sec:optimization_bounds}.

\begin{figure}[h]
	\centering
	\includegraphics[width=0.48\textwidth]{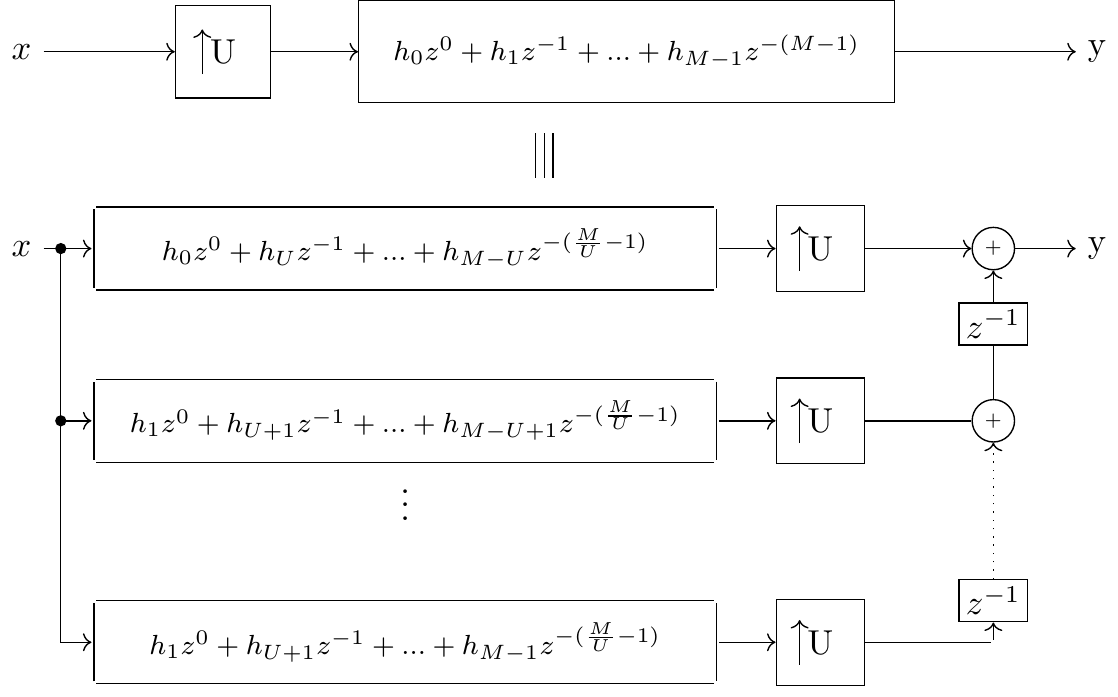}
	\caption{An FIR interpolator and its equilavent polyphase representation.}
	\label{fig:polyphase_direct}
\end{figure}

Polyphase filter banks require preliminary manipulations before using any of these designs. These representations are specifically designed for interpolator or decimator filters. An example FIR interpolator with upsampling ratio $U$ and number of filter coefficients $M$ is in Fig. \ref{fig:polyphase_direct} with its corresponding polyphase representation. Main purpose of polyphase representation is to divide the original filter into subfilters and exploit Noble indetity to reduce the sampling rate of the filter. 

In Fig. \ref{fig:polyphase_direct}, polyphase representation creates a filter bank, which can be implemented using any of the designs in Fig. \ref{fig:coeff_share} \ref{fig:coeff_share_dsp48}, \ref{fig:systolic_coeff_share}, \ref{fig:direct_fir}, \ref{fig:partially_serial_systolic} and \ref{fig:fully_serial_systolic}.

\section{FPGA Synthesis Performance}
\footnote{Reader can regenerate the presented algorithms using codes provided in \cite{arslanFilterCodes}}
In this study Xilinx Zynq UltraScale+ MPSoC ZCU106 is used as the target device. It holds a ZU7EV-2FFVC1156 chip with 230400 lookup tables (LUTs), 460800 registers and 1728 DSP48 elements. This chip is specifically chosen so that synthesis is not bottlenecked by the number of chip resources. Filter coefficients are generated via \textit{randi($\cdot$)} function of MATLAB with random number generated seed \textit{rng(50)}. Generated 0's are simply replaced with -1's.

As stated previously in Sec. \ref{sec:fpga}, comparison is grouped under two main applications, a) FIR filter bank, b) Polyphase filter bank. Filter banks with different number of filters and filter coefficients are implemented using designs in Fig. \ref{fig:coeff_share} \ref{fig:coeff_share_dsp48}, \ref{fig:systolic_coeff_share}, \ref{fig:direct_fir}, \ref{fig:partially_serial_systolic} and \ref{fig:fully_serial_systolic}. For our algorithm, the number of filters in groups is chosen via method described in Sec. \ref{sec:optimization_bounds}. For the polyphase filter, only interpolator structure is implemented since decimator structure is very similar. 

\subsection{FIR filter bank}
Three different FIR filter banks are tested, $K= 8$ and $M=120$ each, $K= 8$ and $M=90$ each, $K= 8$ and $M=60$ each. Number of groups for the sharing algorithm is found via the methods in Sec. \ref{sec:optimization_bounds}. For the partially systolic architecture, filter is accelerated so that only two $M$ length MAC structures are used. The resource performance and the sampling frequency ($F_s$) of the algorithms are in Tables \ref{tab:8_120}, \ref{tab:8_90} and \ref{tab:6_60}.

\begin{table}[h]
\centering
\caption{Performance of FIR Designs for 8 filters with 120 coefficients each.}
\label{tab:8_120}
\begin{adjustbox}{max width=0.48\textwidth}
\begin{tabular}{|c|c|c|c|c|c|c|c|c|}
\hline
\multicolumn{2}{|c|}{}                                                           & Direct FIR &  \begin{tabular}[c]{@{}c@{}}Coeff. \\ Share \end{tabular} & \begin{tabular}[c]{@{}c@{}}Coeff. \\ Share\\ /w DSP48\end{tabular} & \multicolumn{2}{c|}{\begin{tabular}[c]{@{}c@{}}Systolic\\ Coeff. \\ Share\end{tabular}} & Part. Sys & Full. Sys. \\ \hline
\multirow{2}{*}{LUTs}  & \begin{tabular}[c]{@{}c@{}}LUT as \\ Logic\end{tabular} & 1655       & 2518     & 1172                                                        & \multicolumn{2}{c|}{2097}                                                        & 1148      & 960        \\ \cline{2-9} 
                       & \begin{tabular}[c]{@{}c@{}}LUT as\\ Mem.\end{tabular}   & 10049      & 2145     & 2599                                                        & \multicolumn{2}{c|}{1664}                                                        & 10473     & 9964       \\ \hline
\multirow{2}{*}{Regs.} & FFs                                                     & 20284      & 11568    & 12713                                                       & \multicolumn{2}{c|}{8499}                                                        & 9520      & 7282       \\ \cline{2-9} 
                       & Latch                                                   & 0          & 0        & 0                                                           & \multicolumn{2}{c|}{0}                                                           & 0         & 0          \\ \hline
\multicolumn{2}{|c|}{DSP48}                                                      & 960        & 240      & 366                                                         & 79                                      & 32                                     & 240       & 120        \\ \hline
\multicolumn{2}{|c|}{$F_s$}                                                      & 1x         & 1x       & 1x                                                          & 2x                                      & 4x                                     & 4x        & 8x         \\ \hline
\multicolumn{2}{|c|}{Carry8}                                                     & 0          & 192      & 0                                                           & \multicolumn{2}{c|}{0}                                                           & 0         & 0          \\ \hline
\multicolumn{2}{|c|}{Delay}                                                      & 0          & 4      & 0                                                           & \multicolumn{2}{c|}{4}                                                           & 2         & 2          \\ \hline
\end{tabular}
\end{adjustbox}
\end{table}

\begin{table}[h]
\centering
\caption{Performance of FIR Designs for 8 filters with 90 coefficients each.}
\label{tab:8_90}
\begin{adjustbox}{max width=0.48\textwidth}
\begin{tabular}{|c|c|c|c|c|c|c|c|c|}
\hline
\multicolumn{2}{|c|}{}                                                           & Direct FIR &  \begin{tabular}[c]{@{}c@{}}Coeff. \\ Share \end{tabular} & \begin{tabular}[c]{@{}c@{}}Coeff. \\ Share\\ /w DSP48\end{tabular} & \multicolumn{2}{c|}{\begin{tabular}[c]{@{}c@{}}Systolic\\ Coeff. \\ Share\end{tabular}} & Part. Sys & Full. Sys. \\ \hline
\multirow{2}{*}{LUTs}  & \begin{tabular}[c]{@{}c@{}}LUT as \\ Logic\end{tabular} & 1234       & 2215     & 265                                                         & \multicolumn{2}{c|}{1722}                                                        & 863       & 721        \\ \cline{2-9} 
                       & \begin{tabular}[c]{@{}c@{}}LUT as\\ Mem.\end{tabular}   & 6025       & 1542     & 706                                                         & \multicolumn{2}{c|}{1423}                                                        & 6063      & 5718       \\ \hline
\multirow{2}{*}{Regs.} & FFs                                                     & 15149      & 8900     & 6732                                                        & \multicolumn{2}{c|}{7860}                                                        & 7276      & 5572       \\ \cline{2-9} 
                       & Latch                                                   & 0          & 0        & 0                                                           & \multicolumn{2}{c|}{0}                                                           & 0         & 0          \\ \hline
\multicolumn{2}{|c|}{DSP48}                                                      & 720        & 180      & 298                                                         & 68                                      & 32                                     & 180       & 90         \\ \hline
\multicolumn{2}{|c|}{$F_s$}                                                      & 1x         & 1x       & 1x                                                          & 2x                                      & 4x                                     & 4x        & 8x         \\ \hline
\multicolumn{2}{|c|}{Carry8}                                                     & 0          & 192      & 0                                                           & \multicolumn{2}{c|}{0}                                                           & 0         & 0          \\ \hline
\multicolumn{2}{|c|}{Delay}                                                      & 0          & 4      & 0                                                           & \multicolumn{2}{c|}{4}                                                           & 2         & 2          \\ \hline
\end{tabular}
\end{adjustbox}
\end{table}

\begin{table}[h]
\centering
\caption{Performance of FIR Designs for 6 filters with 60 coefficients each.}
\label{tab:6_60}
\begin{adjustbox}{max width=0.48\textwidth}
\begin{tabular}{|c|c|c|c|c|c|c|c|c|}
\hline
\multicolumn{2}{|c|}{}                                                           & Direct FIR &  \begin{tabular}[c]{@{}c@{}}Coeff. \\ Share \end{tabular} & \begin{tabular}[c]{@{}c@{}}Coeff. \\ Share\\ /w DSP48\end{tabular} & \multicolumn{2}{c|}{\begin{tabular}[c]{@{}c@{}}Systolic\\ Coeff. \\ Share\end{tabular}} & Part. Sys & Full. Sys. \\ \hline
\multirow{2}{*}{LUTs}  & \begin{tabular}[c]{@{}c@{}}LUT as \\ Logic\end{tabular} & 699        & 1102     & 204                                                         & \multicolumn{2}{c|}{1301}                                                        & 558       & 479        \\ \cline{2-9} 
                       & \begin{tabular}[c]{@{}c@{}}LUT as\\ Mem.\end{tabular}   & 2448       & 1056     & 407                                                         & \multicolumn{2}{c|}{935}                                                         & 2802      & 2621       \\ \hline
\multirow{2}{*}{Regs.} & FFs                                                     & 7920       & 5805     & 3603                                                        & \multicolumn{2}{c|}{5640}                                                        & 4817      & 3713       \\ \cline{2-9} 
                       & Latch                                                   & 0          & 0        & 0                                                           & \multicolumn{2}{c|}{0}                                                           & 0         & 0          \\ \hline
\multicolumn{2}{|c|}{DSP48}                                                      & 360        & 117      & 159                                                         & 68                                      & 16                                     & 120       & 60         \\ \hline
\multicolumn{2}{|c|}{$F_s$}                                                      & 1x         & 1x       & 1x                                                          & 2x                                      & 3x                                     & 3x        & 6x         \\ \hline
\multicolumn{2}{|c|}{Carry8}                                                     & 0          & 76       & 0                                                           & \multicolumn{2}{c|}{0}                                                           & 0         & 0          \\ \hline
\multicolumn{2}{|c|}{Delay}                                                      & 0          & 3      & 0                                                           & \multicolumn{2}{c|}{3}                                                           & 1         & 1          \\ \hline
\end{tabular}
\end{adjustbox}
\end{table}

From Tables \ref{tab:8_120}, \ref{tab:8_90} and \ref{tab:6_60}, the major advantage of the proposed algorithm is the sampling period of the filter. Without increasing sampling period, proposed algorithm is able to decrease the number of used resources significantly. Coefficient sharing algorithm require more logic resources because it has to implement a summation pyramid at the filter stage as in Fig. \ref{fig:coeff_share}. This is also the same reason why it needs Carry8s. Coefficient sharing algorithm with DSP48 at the filter stage uses much less logic resources in exchange of some DSP48s. 

Systolic Coefficient Sharing algorithm on the other hand is much more complex as per resource usage. For the filter bank with $K= 8$ and $M=120$ case, the subset stage is accelarated twice where as the filter stage is accelerated four times. This translates to 79 of the MACs having twice the sampling frequency of the input data. 79 MACs are in fact the number of MACs used to implement the subset stage. 32 of the MACs require four times the sampling frequency of the ipnut data, which is the number of MACs used to implement the outer summation stage. Thus a total of 111 DSP48s are used in a multirate manner. 

As the number of coefficients decrease, number of DSPs used for the fully systolic design is much less than the systolic coefficient sharing algorithm. This is because the outer summations need a fix number of MACs to compute where as number of MACs needed by fully systolic design is exactly equal to the number of coefficients. This is evident in the $K= 8$ and $M=90$ case. Even though fully systolic approach uses less number of DSP48s, its sampling frequency is much higher compared to the systolic coefficient sharing algorithm. In addition to this, systolic coefficient sharing algorithm uses majority of DSP48s in a much lower sampling frequency. Similar comments can be made for $K= 6$ and $M=60$ case.

Finally, for the cases in Table \ref{tab:8_120} and \ref{tab:8_90}, coefficient sharing with summation pyramid gives the first output 4 clock cycles late than the direct form FIR implementation. This is due to the way the summation pyramid is implemented with delay blocks. Since there are 4 filters in each group, the summation pyramid has 4 stages with a unit delay block at each stage, resulting in 4 delays. Coefficient sharing with systolic implementation has additional delay of 4 clock cycles because of the pipeline delays and upsampling blocks that are implemented via shift registers. Since for the $K= 6$ and $M=60$ case the number of subsets are much smaller, the overall delays for this cases are less than the ones in Table \ref{tab:8_120} and \ref{tab:8_90}. However, any of the proposed coefficient sharing algorithms does not introduce significant delays with respect to the direct form FIR filter.

These synthesis results shows the flexibility and efficiency of the proposed algorithm. It not only reduces the number of operations compared to regular filter banks, but also enables application of other algorithms. This is because, the sharing algorithm is based on regrouping and rewriting the regular convolution equation without changing its base structure.

\subsection{Polyphase Application}
As shown in Fig. \ref{fig:polyphase_direct}, polyphase representation creates a filter bank structure. An interpolator with $U = 2$ upsampling ratio and $M = 60$ coefficient filter, and another interpolator with $U = 3$  and $M = 60$ filter is implemented. Since number of filters in the polyphase representation is determined by the upsampling ratio, partially serial systolic structure is not implemented. The  resource  performance and the sampling frequency ($F_s$) of the algorithms are in Tables \ref{tab:3up_90} and \ref{tab:2up_60}.

\begin{table}[h!]
\centering
\caption{Performance of FIR Designs for an interpolator with 3 upsampling ratio and 90 coefficient filter.}
\label{tab:3up_90}
\begin{adjustbox}{max width=0.48\textwidth}
\begin{tabular}{|c|c|c|c|c|c|c|c|}
\hline
\multicolumn{2}{|c|}{}                                                           & Direct FIR &  \begin{tabular}[c]{@{}c@{}}Coeff. \\ Share \end{tabular} & \begin{tabular}[c]{@{}c@{}}Coeff. \\ Share\\ /w DSP48\end{tabular} & \multicolumn{2}{c|}{\begin{tabular}[c]{@{}c@{}}Systolic\\ Coeff. \\ Share\end{tabular}} & Full. Sys. \\ \hline
\multirow{2}{*}{LUTs}  & \begin{tabular}[c]{@{}c@{}}LUT as \\ Logic\end{tabular} & 267        & 601      & 248                                                         & \multicolumn{2}{c|}{622}                                                         & 284        \\ \cline{2-8} 
                       & \begin{tabular}[c]{@{}c@{}}LUT as\\ Mem.\end{tabular}   & 701        & 244      & 302                                                         & \multicolumn{2}{c|}{266}                                                         & 681        \\ \hline
\multirow{2}{*}{Regs.} & FFs                                                     & 2756       & 1999     & 1943                                                        & \multicolumn{2}{c|}{2371}                                                        & 1956       \\ \cline{2-8} 
                       & Latch                                                   & 0          & 0        & 0                                                           & \multicolumn{2}{c|}{0}                                                           & 0          \\ \hline
\multicolumn{2}{|c|}{DSP48}                                                      & 90         & 30       & 51                                                          & 21                                      & 8                                      & 30         \\ \hline
\multicolumn{2}{|c|}{$F_s$}                                                      & 1x         & 1x       & 1x                                                          & 2x                                      & 3x                                     & 3x         \\ \hline
\multicolumn{2}{|c|}{Carry8}                                                     & 5          & 47       & 6                                                           & \multicolumn{2}{c|}{6}                                                           & 5          \\ \hline
\multicolumn{2}{|c|}{Delay}                                                      & 0          & 9      & 0                                                           & \multicolumn{2}{c|}{12}                                                           & 6         \\ \hline
\end{tabular}
\end{adjustbox}
\end{table}

\begin{table}[h!]
\centering
\caption{Performance of FIR Designs for an interpolator with 2 upsampling ratio and 60 coefficient filter.}
\label{tab:2up_60}
\begin{adjustbox}{max width=0.48\textwidth}
\begin{tabular}{|c|c|c|c|c|c|l|c|}
\hline
\multicolumn{2}{|c|}{}                                                           & Direct FIR &  \begin{tabular}[c]{@{}c@{}}Coeff. \\ Share \end{tabular} & \begin{tabular}[c]{@{}c@{}}Coeff. \\ Share\\ /w DSP48\end{tabular} & \multicolumn{2}{c|}{\begin{tabular}[c]{@{}c@{}}Systolic\\ Coeff. \\ Share\end{tabular}} & Full. Sys. \\ \hline
\multirow{2}{*}{LUTs}  & \begin{tabular}[c]{@{}c@{}}LUT as \\ Logic\end{tabular} & 213        & 375      & 205                                                         & \multicolumn{2}{c|}{403}                                                         & 237        \\ \cline{2-8} 
                       & \begin{tabular}[c]{@{}c@{}}LUT as\\ Mem.\end{tabular}   & 642        & 310      & 324                                                         & \multicolumn{2}{c|}{222}                                                         & 680        \\ \hline
\multirow{2}{*}{Regs.} & FFs                                                     & 2197       & 1744     & 1727                                                        & \multicolumn{2}{c|}{1606}                                                        & 1829       \\ \cline{2-8} 
                       & Latch                                                   & 0          & 0        & 0                                                           & \multicolumn{2}{c|}{0}                                                           & 0          \\ \hline
\multicolumn{2}{|c|}{DSP48}                                                      & 58         & 30       & 37                                                          & \multicolumn{2}{c|}{21}                                                          & 30         \\ \hline
\multicolumn{2}{|c|}{$F_s$}                                                      & 1x         & 1x       & 1x                                                          & \multicolumn{2}{c|}{2x}                                                          & 2x         \\ \hline
\multicolumn{2}{|c|}{Carry8}                                                     & 2          & 14       & 3                                                           & \multicolumn{2}{c|}{3}                                                           & 2          \\ \hline
\multicolumn{2}{|c|}{Delay}                                                      & 0          & 4      & 0                                                           & \multicolumn{2}{c|}{8}                                                           & 4         \\ \hline
\end{tabular}
\end{adjustbox}
\end{table}

From Tables \ref{tab:3up_90} and \ref{tab:2up_60}, proposed algorithm is again very efficient. Compared to the direct FIR filter, it uses much less FFs and memory LUTs. Proposed algorithm with filter blocks implemented with summation pyramid structure is especially efficient resource wise. Compared to the systolic architecture, propose algorithm uses comparable number of resource but requires much less sampling frequency. Carry8s are used in all implementations since polyphase structure requires summation operations at the output. Base architecture of the proposed algorithm naturally needs more Carry8s due to the summation implementation at the filter stage. 

Systolic implementation of coefficient sharing algorithm is again as efficient as fully systolic implementation, resource wise. For the $U=3$ and $M = 90$ case, systolic coefficient sharing has 21 of its DSP48s working at twice the sampling frequency of the input, compared to the 30 of the DSP48s of fully systolic implementation working at three times the sampling frequency of the input. This translates to less power usage compared to the fully systolic approach. Hence proposed algorithm is overall less demanding compared to other filter design architectures overall.

Similar to the regular filter bank implementation, coefficient sharing with summation pyramid gives the first output 9 clock cycles late than the direct form FIR implementation for the interpolator with $U = 3$ and $M = 90$. Coefficient sharing with systolic implementation is on the other hand is the slowest. In addition to the delays in the summation pyramid design, overall data integrity pipelines are the main reasons for increased delay.

\section{Conclusion}
\label{sec:conclusion}
In this paper, a new coefficient sharing algorithm for parallel filter banks, that shares the coefficients between a number of filters is presented. Filter banks are widely used in radar signal processing and signal synchronization in communications field and quickly becomes one of the most resource heavy blocks in an FPGA for long sequences and high number of filters. The algorithm groups the filters and finds the similarities within the specified group. The coefficients are rearranged according to the rules discussed in Section \ref{sec:coeff_share}. 

The optimization bounds of the algorithm clearly shows the efficiency of the algorithm with respect to filter length and number of filters in a group, giving an early idea on how to group the filters in a filter bank in order to achieve better resource usage performance. 

Finally the post-synthesis results in an FPGA shows the resource performance with respect to regular filter banks and commonly used systolic architectures. Compared to direct form FIR structure, coefficient sharing algorithm has up to \% 50 less resource usage while working at the same sampling frequency. Systolic designs have comparable resource usage however they also require much higher sampling rate, depending on the sharing coefficient. Coefficient sharing algorithm in which filter stage implemented with multiply-and-accumulate increases the efficiency of the approach further. Especially the DSP48 efficiency of the algorithm is clear.

Algorithm presented in this paper is scalable and flexible. Coefficient sharing algorithm simply implements the same filter bank in an elaborate manner, hence enabling usage of other coefficient sharing algorithms. An example of systolic implementation of the coefficient sharing algorithm is presented in this paper, which provides less number of DSP48s used in lesser sampling frequency compared to regular systolic approaches. A further reduction of resource usage is possible using other well known coefficient sharing algorithms that optimizes single filters such as in \cite{fliege1994multirate,jordan1986correlation,karlsson2005implementation}.

\bibliographystyle{IEEEtran}
\bibliography{references}

\clearpage

\end{document}